\documentclass[twocolumn,floatfix,showpacs]{revtex4}
\usepackage{graphicx}
\usepackage{amsmath}
\usepackage{revsymb}
\usepackage{bm}

\begin{document}

\title{Momentum-dependent light scattering in a 2D Heisenberg antiferromagnet }

\author{A. Donkov and A. V. Chubukov}
\affiliation{
Department of Physics, University of Wisconsin, Madison, WI 53706, USA}
\date{\today}
\begin{abstract}
Motivated by the achievements of the $X$-ray scattering technique, we
 analyzed the profile of the light scattering intensity $R(q, \omega)$ 
at a finite $q$ in a 2D Heisenberg antiferromagnet. 
Previous Raman scattering studies at $q=0$ identified the 
two-magnon peak in $B_{1g}$ scattering geometry. 
We found that the $B_{1g}$ peak disperses downwards at a finite $q$, and its
 intensity increases, reaching its maximum at $q= q_0 = (0, \pi)$ and symmetry
 related points. In addition, the intensity in the $A_{1g}$ geometry 
 becomes non-zero  at a finite $q$, and also displays a two-magnon peak which gains strength and 
disperses to larger frequencies with increasing $q$,
 and reaches its highest intensity at $q_0$. We found that the profile of  $R(q_0, \omega)$ is equivalent in $A_{1g}$ and $B_{1g}$ geometries. 

\end{abstract}


\maketitle

Raman scattering is a powerful probe of magnetic and electronic correlations in interacting electron systems. Raman studies of  antiferromagnetic 
parent compounds of the  cuprates in early 90's
provided the first estimates of the Heisenberg exchange integrals
 in $La_2CuO_4$ and $YB_2C_3O_6$. The values of 
$J \sim 120-140 meV$  were extracted from the positions of the 
 $B_{1g}$ two-magnon peak which, according to the theory, are
 located at approximately $2.7 J$~\cite{review}.  The values of $J$ extracted from the Raman studies were later confirmed by  neutron scattering measurements~\cite{coldea}. 

\begin{figure}[htb!]
\includegraphics{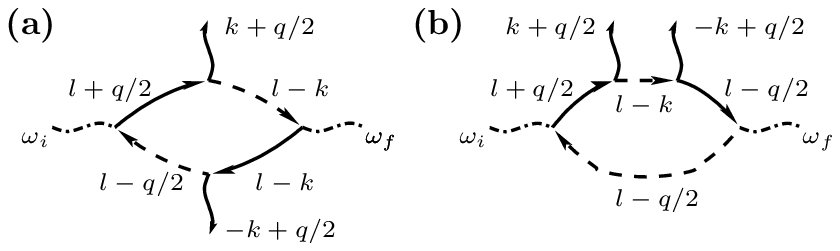} 
\caption{ The diagrams for the Raman matrix element $M_R (k, q)$ 
Solid and dashed lines represent conduction and valence electrons. 
The dash-dotted lines represent incoming and outgoing photons, and 
the solid wavy lines represent magnons. The full set includes an extra diagram (not shown) which cancels out a parasitic contribution from the diagram b, 
see~\protect\cite{ChubukovFrenkel95}.}
\label{fig:1}
\end{figure}

 The traditional framework
 for the understanding of the two-magnon Raman scattering in antiferromagnets has been the semi-phenomenological Loudon-Fleury model 
for the interaction of light with spin degrees of
 freedom~\cite{LF}. The model assumes the existence of the 
matrix element $M_R$ for the
 direct process in which incident photon with energy $\omega_i$ comes in, 
an outgoing photon with energy $\omega_f$ comes out, and  
two magnons are excited with energies $\Omega(k_1)$ and $\Omega (k_2)$
 (see Fig. \ref{fig:1}). 
The energy conservation implies 
$\Omega (k_1) + \Omega (k_2) = \hbar (\omega_i-\omega_f) = \hbar \omega$. 
 The transferred frequency $\omega$ is usually order of magnitude smaller than $\omega_{i,f}$.  

 Quite generally, the matrix element $M_R$ depends on the
 two frequencies $\omega_i$ and $\omega_f$,  the 
two momenta, either of two 
photons or two magnons, and on the polarizations $\mathbf{\hat e}_i$ 
and $\mathbf{\hat e}_f$ of the incoming and outgoing light.  For the two 
 most frequently studied cases of $A_{1g}$ and $B_{1g}$ scattering, the 
 polarizations are 
$\mathbf{\hat e}_i = (\mathbf{\hat x}+\mathbf{\hat y}) / \sqrt{2}, 
\mathbf{\hat e}_f = (\mathbf{\hat x}+\mathbf{\hat y}) / \sqrt{2},$ 
for  $A_{1g}$ scattering, and  $ \mathbf{\hat e}_i = (\mathbf{\hat x}+\mathbf{\hat y}) / \sqrt{2}, 
\mathbf{\hat e}_f = (\mathbf{\hat x}-\mathbf{ \hat y}) / \sqrt{2},$ for
 B$_{1g}$ scattering. 
For non-resonant scattering, which we consider, the
frequency dependence of $M_R$ only affects the overall factor, and the interesting physics comes from the momentum dependence of $M_R$.
 The two magnon momenta can be parameterized as $k_1 = k +q/2$, $k_2 = -k + q/2$, where $q$ is the transferred photon momentum, then $M_R = M_R (k,q)$.  At $q=0$, $M_R \propto \cos k_x + \cos k_y$ in $A_{1g}$ geometry, and $M_R \propto \cos k_x - \cos k_y$ in $B_{1g}$ geometry.    

Without the final state interaction,  the  intensity of the absorption of light 
$R_0(q, \omega)$   
 is obtained from the Fermi golden rule, and  is given by 
\begin{eqnarray}
R_0 (q, \omega) &\propto& \sum_k M^2_R (k,q) \times \nonumber\\
 &\times& \delta \left[\hbar \omega -(\Omega (k+q/2) + \Omega (-k+q/2)\right].
\label{1}
\end{eqnarray}   
Below we label by $R (q, \omega)$ the full intensity, which includes
the final state interaction. 

In Raman scattering experiments with visible light,
 typical frequencies of the incident light are $\omega_{i,f} 
/ (2 \pi) \sim 5  10^{14}$ Hz, and the corresponding 
 wave-vectors $ q_{i,f} \sim 1.5  10^{7}$ m$^{-1}.$
 Typical momenta $k$ for  magnons near magnetic Brillouin zone boundary, 
which mostly contribute to Raman scattering,
  are of the order $ \pi/d, $ where $d \sim 10^{-10}$m is the  lattice constant.
Accordingly,  $ k \sim  10^{10}$ m$^{-1}$, three orders of magnitude larger than $q_{i,f}$, in which case one
 can safely set $q=0$ in Eq.~(\ref{1})~\cite{HayesLoudon78}.
  The approximation $R(q, \omega) \approx R (q=0, \omega)$ 
has been used in all previous calculations of the Raman intensity~\cite{shastry,devereaux,ChubukovFrenkel95}.

In recent years, however, a new resonant inelastic X-ray scattering technique (RIXT) has been developed, which allows one to probe the intensity $R(q, \omega)$ 
at a finite $q$~\cite{rixsreview,vdBrink,HillBlumberg}. The frequencies of $X$ rays 
are tuned to the energies of atomic transitions  and are
 several orders of magnitude larger than for conventional Raman experiments:
 $\omega_i / (2 \pi) \sim 2 10^{18} $ Hz,~\cite{NADyson90}. 
The corresponding momenta are then $ q_{i,f} \sim 6  10^{10}$ m$^{-1}$. These momenta  are comparable to the typical magnon momenta $k$.
As a result, RIXT technique  allows one to measure the light intensity $R(q, \omega)$ 
at a finite $q$.

A RIXT study of the undoped and weakly doped $La_{2-x}Sr_xCuO_4$ and undoped 
$Nd_2CuO_4$ 
has recently been carried out by Hill et al~\cite{HillBlumberg}. In both undoped compounds,  the authors scanned various momenta $q$, 
 and at $q = q_0 = (0,\pi)$ observed a sharp peak in $R(q_0, \omega)$ 
at $500 meV$. On deviations from $q_0$, the peak moves to smaller frequencies and broadens.  The doping dependence of the 
peak  at $q_0$ is similar to that of the two-magnon Raman peak -- both 
 broaden with doping and disappear near optimal doping. Based on this similarity, the authors of ~\cite{HillBlumberg} argued that the peak at $q_0$ is likely a
 two-magnon feature at a finite $q$. This is corroborated by the fact that 
the density of states (DOS) of two non-interacting magnons has a peak at $(0,\pi)$ at $3.5 J \sim 500 meV$. 

The presence of the peak in the DOS, however, does not necessary imply the 
 peak in $R_0 (q_0, \omega)$ as the matrix element $M_R$ may counterweight the DOS effect. 
Besides, magnons do interact in the final state, and this interaction 
 may substantially affect the profile of $R(q, \omega)$ compared to that of $R_0 (q, \omega)$. 

In this communication, we analyze  $R(q, \omega)$ in $A_{1g}$ and $B_{1g}$ geometries 
 in two-dimensional (2D) 
antiferromagnetic Mott insulators (e.g., parent compounds of high-temperature superconductors). Like we said above, at $q=0$,  the Raman response in the $B_{1g}$ geometry 
 contains the   two-magnon peak at $\omega \approx 2.7 J$.
In  $A_{1g}$ geometry the Raman intensity vanishes at $q=0$ 
 because  Loudon-Fleury and Heisenberg Hamiltonians 
 commute~\cite{shastry,ChubukovFrenkel95}. 
We consider how both features are modified at a finite $q$. We show that the position  of the two-magnon peak in the $B_{1g}$ geometry evolves with $q$, and
 at $q_0$ the peak is located at $2.4 J \sim 340 meV$.
 We show that the intensity of the $B_{1g}$ peak actually increases at a finite $q$ and is the     
 largest at $q_0$ and symmetry-related points.  We further
 show that in $A_{1g}$ geometry, 
$R(q, \omega)$ is non-zero at a finite $q$, and the intensity is again
 the largest at $q_0$. We found that for $q = q_0$, the intensities 
 $R (q, \omega)$ in $A_{1g}$ and $B_{1g}$ channels are equivalent both with and without final state interaction.  

Our results cannot be directly applied to the experiments by Hill et al~\cite{HillBlumberg} as they found the peak at $q_0$ in the geometry when
 the incident polarization is along $c-$axis. To obtain the matrix element
 in such geometry one would need to consider hopping between 2D planes. 
 At the same time, our $A_{1g}$  and $B_{1g}$ results show that the profile of the peak in $R(q_0, \omega)$ is predominantly determined by the interplay between  the two-magnon DOS and magnon-magnon interaction, while the momentum dependence of the matrix element $M_R$ does not play a substantial role near $q_0$. We therefore expect that the (identical) profile of $R(q_0, \omega)$ in $A_{1g}$ and $B_{1g}$ geometry is similar to that for $c-$axis polarization of the light, and
the intensity is just stronger in $c-$axis geometry.

The derivation of the matrix element $M_R$ for $x-$ray scattering 
 which involves high-energy photons in general requires a 
 careful consideration of  core atomic transitions
~\cite{HillBlumberg,platzman}.  The outcome
 of this consideration is some effective direct coupling between $x-$rays and two magnons.  In the following, we adopt an approach borrowed from Raman studies and 
derive the Raman matrix element by the same method as in Ref.~\cite{ChubukovFrenkel95}, which we extended to finite $q$.  Namely, we depart from 
the 2D half-filled 
Hubbard model with strong on-site repulsion $U$ and the nearest-neighbor hopping $t$,
  expand the hopping term
 in  vector potential ${\bf A}$, and restrict
 with the on-site interaction term ${\cal H}_{int} = -(e/\hbar c) \sum_q {\bf j}_q {\bf A}_{-q}$
 between the vector potential and the fermionic current. We then
 introduce antiferromagnetic long-range order parameter 
$\Delta$ ($ \approx U/2$ 
 at $U >> t$ which we only consider),
 and split quasiparticles into conduction and valence bands, with energies $\pm \sqrt{\Delta^2 + \epsilon^2_k}$, where $\epsilon_k = -2t (\cos k_x + \cos k_y)$.We then obtain from the model the 
 interaction  vertices between magnons and conduction and valence fermions. At large $U/t$, the dominant interactions are the ones in which fermion flips from the conduction to the valence band and vise versa after emitting a magnon. To keep calculations under control, we extended the model to ``large S'' by introducing $2S$ fermionic flavors~\cite{largeS}. For large $S$, the corrections to our
 vertices are small, of order $1/S$. 

Like we just said, for energies of the incident phonons well above Hubbard $U$, the matrix element $M_R$ comes from the core level atomic transitions rather then from the excitations across the Hubbard gap. Still, we believe that the derivation using ``Raman technique'' is useful as it shows
 the difference between the conventional Raman scattering with $q=0$ and a finite $q$ scattering in different scattering geometries. Besides, as we will see, for scattering near $(0,\pi)$ the profile of $R(q, \omega)$ is determined predominantly by the structure of the DOS and magnon-magnon interaction, while the 
 matrix element $M_R$ can be safely set to be a constant.  This implies that the results below could be obtained with just a phenomenological $M_R$. 

The processes which contribute to the matrix element $M_R$ in finite $q$ Raman 
 scattering are the ones in which incident light creates a particle-hole pair, which creates two magnons,
 either from two different lines in the particle-hole bubble, or from the same line, and then 
annihilates into outgoing light.
  The corresponding two diagrams are presented in Fig.~\ref{fig:1}. [The diagram 
b has a ``parasitic'' contribution which for $q=0$ is canceled out by the additional diagram with an extra fermion-fermion interaction~\cite{ChubukovFrenkel95}.  We verified 
 that  the same cancellation occurs also at a finite $q$]. Evaluating the two diagrams and adding them up, we obtain the matrix element in various geometries  in the form 
\[ M_R(k,q)= -8 t^2 \left[ \frac{2 \Delta}{4 \Delta^2-\omega_i^2} \right] \times \]
\[ \{ [\mathbf{\hat e}_{ix} \mathbf{\hat e}^*_{fx} \cos(q_x/2) + \mathbf{\hat e}_{iy} \mathbf{\hat e}^*_{fy} \cos(q_y/2)](\lambda_{k+q/2} \mu_{-k+q/2} + \]
\[ + \lambda_{-k+q/2} \mu_{k+q/2}) - [\mathbf{\hat e}_{ix} \mathbf{\hat e}^*_{fx} \cos(k_x) + \mathbf{\hat e}_{iy} \mathbf{\hat e}^*_{fy} \cos(k_y) ] \times \]
\begin{equation}
\times(\mu_{k+q/2} \mu_{-k+q/2} + \lambda_{k+q/2} \lambda_{-k+q/2}) \}, 
\label{eqn:MR}
\end{equation} 
where 
 $\mu_k = \frac{1}{\sqrt{2}} \sqrt{\frac{1}{\sqrt{1-\gamma_k^2}} + 1}, ~
 \lambda_k= \frac{\gamma_k}{|\gamma_k| \sqrt{2}} \sqrt{\frac{1}{\sqrt{1-\gamma_k^2}} - 1}$, and $ \gamma_k = \left(\cos(k_x) + \cos(k_y)\right)/2$.
The magnon energy  $\Omega_{k}$  is given by   
$\Omega_k = E_{max} \sqrt{1-\gamma_k^2},$ where $E_{max} =4 J S (=2J)$, and $J = t^2/4U$.
At $q=0$, the expression in Eq.~(\ref{eqn:MR}) coincides with that in earlier works~\cite{ChubukovFrenkel95,shastry}. 
 In particular,
one can easily verify that $M_R (k,0)$ vanishes in $A_{1g}$ geometry.  At $q \neq 0$, the matrix element is non-zero in all scattering geometries, including $A_{1g}$.  Furthermore, a simple  trigonometric analysis shows that 
 at  $q =q_0 = (0,\pi)$, the values of $M_R (k,q_0)$ in $B_{1g}$ and $A_{1g}$ geometries are identical.

\begin{figure}[htb!]
\includegraphics{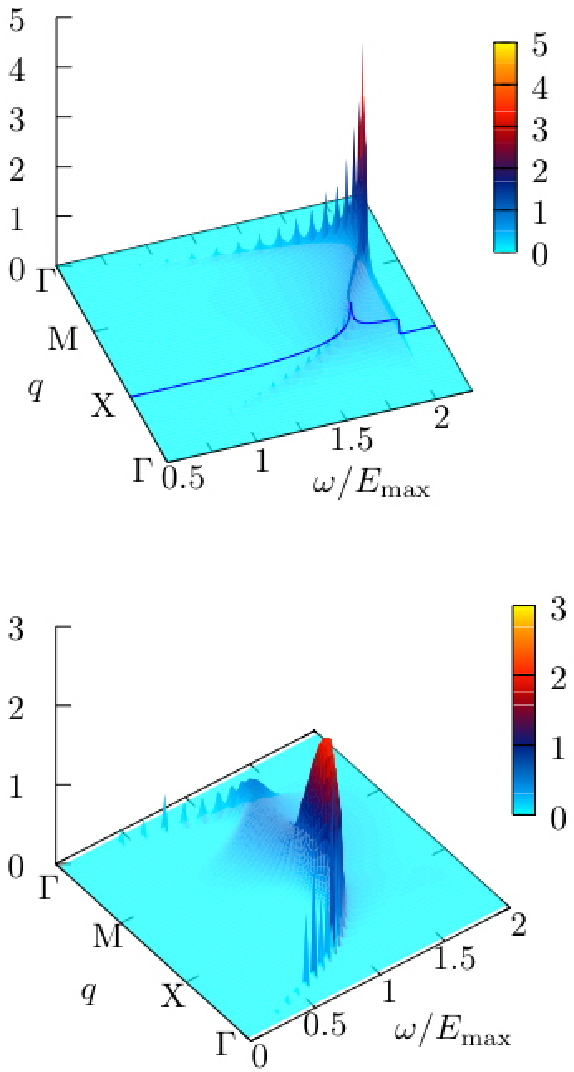} 
\caption{The scattering intensity in the $A_{1g}$ geometry along 
$\Gamma-X-M-\Gamma$ direction, where $\Gamma = (0,0)$, $X = (\pi,\pi)$, and 
$M = (0, \pi)$. 
 Top panel, without final state magnon-magnon interaction. Bottom panel - with the final state 
 interaction. The largest intensity for the full $R(q, \omega)$ 
is at $q_0 = (0, \pi)$.}
\label{fig:2}
\end{figure}

\begin{figure}[htb!]
\includegraphics{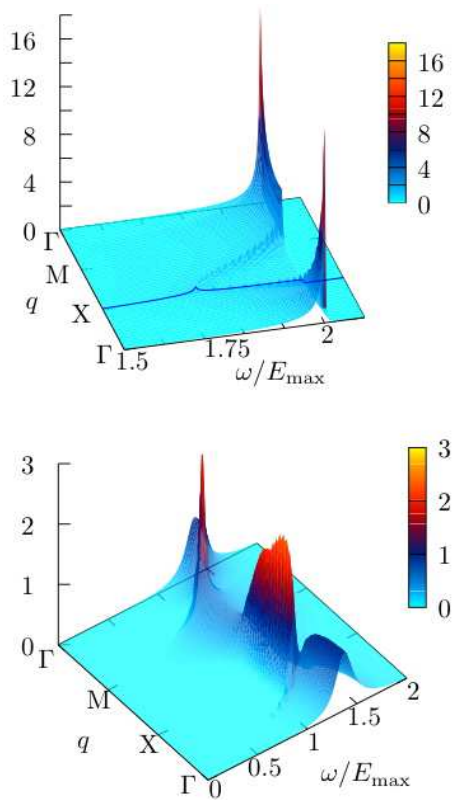} 
\caption{Same as in Fig.\protect\ref{fig:2}, but for $B_{1g}$ geometry.
The intensity for the full $R(q, \omega)$ is again the largest for $q_0 = (0,\pi)$.}
\label{fig:3}
\end{figure}

\begin{figure}[htb!]
\includegraphics{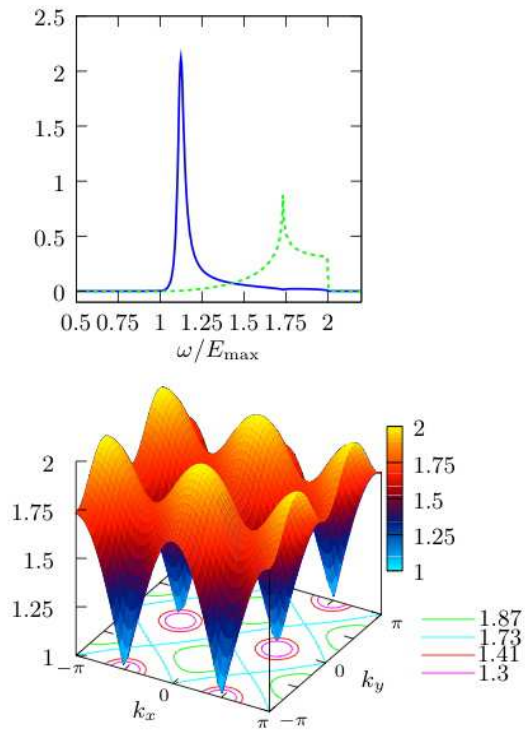} 
\caption{Top panel: the scattering intensities at $q_0 = (0, \pi)$ 
with and without final state interaction (solid and dashed lines, respectively). Note that due to  a final state interaction,  the two-magnon peak gets sharper, becomes more symmetric and shifts to a lower frequency. Bottom panel: 2D constant energy contours for the total  energy of two magnons $\Omega (k+q_0/2) + \Omega (-k+q_0/2)$. The extended van-Hove singularity is at $\Omega (k+q_0/2) + \Omega (-k+q_0/2) = 2 \sqrt{3} J$.}
\label{fig:4}
\end{figure}

Without final state interaction, the scattering intensity $R_0 (q, \omega)$ 
 is given by Eq.~(\ref{1}). In the top panels of Fig. \ref{fig:2}, 
\ref{fig:3},
 we present the results for $R_0 (q, \omega)$  for $A_{1g}$ and $B_{1g}$ scattering.  We see that in $A_{1g}$ geometry, the two-magnon peak gradually emerges
 as $q$ increases, and  progressively shifts to  
larger frequencies. In $B_{1g}$ geometry, the largest intensity without final state interaction is still at $q=0$. At a finite $q$, the 
peak position shifts to lower frequencies. 
 In Fig.~\ref{fig:4}, top panel,
 we present the profile of $R_0 (q_0, \omega)$ (a dashed line) which, we remind, is identical in $A_{1g}$ and $B_{1g}$ geometries.  We see that $R_0 (q_0, \omega)$ has  a sharp peak at $\omega = \sqrt{3} E_{max}$
 followed by the shoulder at $\omega = 2 E_{max}$, which is the maximum value of the energy of two magnons. The position of the peak at $\sqrt{3} E_{max}$ can 
 be traced back to the behavior of $\delta [ \omega - (\Omega (k+q_0/2) + \Omega (k-q_0/2))]$, as $\Omega (k+q_0/2) + \Omega (k-q_0/2)$  has an extended 
 van-Hove singularity at the set of $k-$points for which 
 $\Omega (k_0+q_0/2) + \Omega (k_0-q_0/2) = \sqrt{3} E_{max}$ (see the bottom panel of Fig.~\ref{fig:4}).  The Raman matrix element $M(k, q_0)$ is regular 
 and we found that its $k-$dependence 
only weakly  affects the form   of $R_0(q_0, \omega)$.

The Fermi golden-rule results for $R_0(q, \omega)$ are of limited use, however,
 as 
 two excited magnons interact in the final state. The magnon-magnon interaction  is not small for  $S=1/2$, and in general substantially affects the scatting profile.  We obtained the vertex for  magnon-magnon interaction by 
 a standard technique: we used the fact that the half-filled Hubbard model at large $U$ reduces to the 
 Heisenberg  2D antiferromagnet  with nearest 
neighbor exchange, $H= J \sum_{<i,j>} S_i\cdot S_j$, re-expressed spin operators in terms of Holstein-Primakoff bosons $\alpha$ and $\beta$ which describe 
excitations in the two sublattices, 
and restricted with  only the interaction term $\alpha^\dagger \beta^\dagger \alpha \beta$ which describes multiple scattering of two excited magnons.
 This procedure is similar to the one used in previous works (see, e.g., 
~\cite{CanaliGirvin92,ChubukovFrenkel95}), the new element is that now the total momentum of the two magnons is non-zero.  The relevant 
interaction part of the Heisenberg Hamiltonian is 
\[ 
H' = \sum_k{}^{'} \sum_l{}^{'}\Gamma_q(k,l)\alpha^\dagger_{k+q/2} \beta^\dagger_{-k+q/2} \beta_{-l+q/2} \alpha_{l+q/2}, 
\]
where the primes indicate that the summation is over the magnetic Brillouin zone, and 
\begin{eqnarray}
\lefteqn{\Gamma_q(k,l) = -\frac{8 J}{N} \times \left\{ \right.} \hspace{-0.1in} \nonumber\\
&& \gamma_{k-l} \left( \mu_{k+q/2} \mu_{l+q/2} \mu_{-k+q/2} \mu_{-l+q/2} +(\mu\leftrightarrow \lambda) \right)\nonumber\\
&&\mbox{} + \gamma_{q} \left( \mu_{k+q/2} \lambda_{l+q/2} \lambda_{-k+q/2} \mu_{-l+q/2} + (\mu\leftrightarrow \lambda) \right) \nonumber \\
&&\!\!\!\!\!\!\!\mbox{} -  \frac{1}{2} \left[ \right.
\gamma_{-k+q/2} \left( \mu_{k+q/2} \mu_{l+q/2} \mu_{-k+q/2} \lambda_{-l+q/2} + (\mu\leftrightarrow \lambda) \right) \nonumber \\
&&\;\;\mbox{} + \gamma_{-l+q/2} \left( \mu_{k+q/2} \mu_{l+q/2} \lambda_{-k+q/2} \mu_{-l+q/2} + (\mu\leftrightarrow \lambda) \right) \nonumber \\
&& \;\;\mbox{} +\gamma_{k+q/2} \left( \mu_{k+q/2} \lambda_{l+q/2} \mu_{-k+q/2} \mu_{-l+q/2} + (\mu\leftrightarrow \lambda) \right) \nonumber \\
&& + \left. \left. \! \!  \gamma_{l+q/2} \left( \lambda_{k+q/2} \mu_{l+q/2} \mu_{-k+q/2} \mu_{-l+q/2} + (\mu\leftrightarrow \lambda)\right) 
\right]
\right\}. \nonumber
\label{eqn:Gammakl}
\end{eqnarray}
We computed the renormalization due to magnon-magnon interaction in the RPA approximation, by projecting the interaction onto $M_R (k, q)$ for different geometries, i.e., by approximating $\Gamma_q(k,l)$ by $B_q M_R(k,q) M_R(l,q)$, where
 $B_q =  \sum'_{k,l} \Gamma_q(k,l)  M_R(k,q)  M_R(l,q)/(\sum'_k M^2_R(k,q))^2$.  In this approximation,  the total scattering intensity 
\begin{equation}
R(q, \omega)  \propto  Im \frac{R_0(q,\omega)}{1 - B R_0 (q, \omega)},
\label{a_1}
\end{equation}
where 
\[
R_0(q,\omega) =  \sum_k{}^{'} \frac{ |M_R (k,q)|^2}{\omega-(\Omega_{k+q/2}+\Omega_{-k+q/2})+i \delta}.
\]
We set $S=1/2$ at the end of calculations. For $q$ near $q_0 = (0, \pi)$, we 
 obtained $B \approx -3$, i.e, the
 final state interaction is quite important.

The results for the scattering intensity $R (q, \omega)$ renormalized by the 
final state magnon-magnon interaction are presented in the bottom panels of 
Fig.~\ref{fig:2}, \ref{fig:3}.   
We see that both $A_{1g}$ and $B_{1g}$ scattering are substantially  sharpened up around $q =q_0$, where the intensity is still at maximum in both geometries. 

We verified that the final state interaction does not break the 
 equivalence between the scattering intensities in $A_{1g}$ and $B_{1g}$ geometries right at $q_0$. In Fig.~\ref{fig:4}, top panel, we plot the full 
intensity 
$R(q_0, \omega)$ versus frequency (a solid line).  Comparing this result with 
$R_0 (q, \omega)$ (a dashed line on the same figure),
  we see that after the final state interaction, the peak gets  sharper and more symmetric, and also shifts to a lower frequency, about $1.2 E_{max}$.
{\it That the fully renormalized intensity} $R (q, \omega)$ {\it has a  sharp 
maximum at $q_0$ is the central result  of our paper}. 

The final state magnon-magnon interaction at a finite $q$ was earlier analyzed
 by Lorenzana and Sawatzky~\cite{lorenz} in the study of phonon-assisted optical absorption.  In their case, a finite momentum $q$ of two magnons is 
equivalent to a phonon momentum. Their effective matrix element for two-magnon vertex is different from ours and belongs to two-dimensional representation of the tetragonal group 
$D_{4h}$.  Still, our results are very  similar to theirs -- they also found 
 that magnon-magnon interaction gives rise to a sharp peak in the absorption near $1.2 E_{max}$.   The only distinction between our results and their is that in their case the peak moves to higher frequencies on moving from $q_0$ towards $(0,0)$, while we found that $A_{1g}$ peak moves to smaller frequencies, and $B_{1g}$ peak position remains almost unchanged   

Our results for the full $R(q, \omega)$ partly agree and partly disagree with the experimental data for  the 
 inelastic resonance $X-$ray scattering of
 antiferromagnetic parent compounds $La_2CuO_4$ and $Nd_2CuO_4$
~\cite{HillBlumberg}.   On one hand, the  position {\it and} the profile of the peak fully  agrees with our 
$R(q, \omega)$. Namely, the largest peak is at $q= q_0$, and it moves to smaller frequency and broadens on deviations from $q_0$.
 On the other hand,  we found, in agreement with~\cite{lorenz}  
that due to final state interaction, the peak in 
$R(q_0, \omega)$ shifts down to $1.2 E_{max} =2.4 J\sim 340 meV$.
  This frequency is smaller than the experimental peak frequency 
of  $0.5 eV \approx 3.6J$, which is close to the peak position at $q_0$ in the absence of the final state interaction. The peak shifts to a higher frequency if we include   quantum corrections to the magnon dispersion~\cite{oguchi}, neglected in our analysis. Another possible explanation of the discrepancy may be that  the RPA-type analysis of magnon-magnon interaction overestimates the 
 effect of the final state interaction.  For $c-$axis polarization of light, as in ~\cite{HillBlumberg}, the final state interaction involves the exchange between layers and generally should lead to a smaller down-shift of the peak position at $(0,\pi)$ from that fore non-interacting magnons~\cite{blu_pr}. 
On the other hand, we didn't find the sharpest peak at $(0,\pi)$ for non-interacting magnons, e.g., in $B_{1g}$ geometry, the peak at $(0,0)$ is the sharpest. 
Magnon-magnon interaction substantially reduces this last peak, and the one at $(0,\pi)$ becomes the largest. However, this only happens if magnon-magnon interaction is strong enough.

To summarize, we analyzed the profile of the X-ray scattering intensity $R(q, \omega)$ at finite $q$ in a 2D Heisenberg antiferromagnet.  Previous Raman scattering studies at $q=0$ identified the two-magnon peak in $B_{1g}$ geometry. 
We found that the $B_{1g}$ peak slightly 
disperses downwards at a finite $q$, and its
 intensity increases, reaching its maximum at $q= q_0 = (0, \pi)$ and symmetry
 related points. Simultaneously, at a finite $q$, the intensity in $A_{1g}$ geometry also becomes non-zero, and the $A_{1g}$ Raman profile displays 
 a two-magnon peak which disperses to
 larger frequencies with increasing $q$ and reaches its highest intensity at $q_0$. We found that the profile of  $R(q_0, \omega)$ is equivalent in $A_{1g}$ and $B_{1g}$ geometries. 

We thank G. Blumberg for useful conversations, constructive criticism,
 and for sharing Ref.~\cite{HillBlumberg} with us. 
The research was supported by  NSF DMR 0240238.

\end{document}